# Title: Coexistence of electron whispering-gallery modes and atomic collapse states in graphene/WSe$_2$ heterostructure quantum dots

**Authors:** Qi Zheng[1, §], Yu-Chen Zhuang[2, §], Qing-Feng Sun[2,3,4,†], Lin He[1,†]

**Affiliations:**
[1] Center for Advanced Quantum Studies, Department of Physics, Beijing Normal University, Beijing, 100875, People's Republic of China
[2] International Center for Quantum Materials, School of Physics, Peking University, Beijing, 100871, China
[3] Collaborative Innovation Center of Quantum Matter, Beijing 100871, China
[4] Beijing Academy of Quantum Information Sciences, West Bld. #3, No. 10 Xibeiwang East Road, Haidian District, Beijing 100193, China

[§]These authors contributed equally to this work.
[†]Correspondence and requests for materials should be addressed to Qing-Feng Sun (email: sunqf@pku.edu.cn) and Lin He (e-mail: helin@bnu.edu.cn).

The relativistic massless charge carriers with a Fermi velocity of about $c/300$ in graphene enable us to realize two distinct types of resonances ($c$, the speed of light in vacuum). One is electron whispering-gallery mode in graphene quantum dots arising from **the Klein tunneling of the massless Dirac fermions. The other is** atomic collapse state, which has never been observed in experiment with real atoms due to the difficulty of producing heavy nuclei with charge $Z > 170$, however, can be realized near a Coulomb impurity in graphene with a charge $Z \geq 1$ because of the "small" velocity of the Dirac excitations. Here, unexpectedly, we demonstrate that both the electron whispering-gallery modes and atomic collapse states coexist in graphene/WSe$_2$ heterostructure quantum dots due to the Coulomb-like potential near their edges. By applying a perpendicular magnetic field, evolution from the atomic collapse states to unusual Landau levels in the collapse regime are explored for the first time.

Many exotic electronic properties of graphene are rooted in its relativistic massless charge carriers (*1-4*). For example, the massless Dirac fermions nature of the charge carriers in graphene enables us to demonstrate several oddball predictions by quantum electrodynamics (QED), among which the Klein tunneling (*5*) and atomic collapse (*6-8*) are the two most famous effects that have attracted much attention. Very recently, it was demonstrated that the two effects lead to the formation of two types of quasibound states in graphene (*9-18*). The Klein tunneling, *i.e.*, the anisotropic transmission of the massless Dirac fermions across the potential barrier, in graphene leads to the formation of quasibound states in circular *p-n* junctions, *i.e.*, graphene quantum dots (GQDs), via whispering-gallery modes (WGMs) (*9-15*). Because of the "small" velocity of the Dirac fermions, a Coulomb impurity in graphene with a charge $Z \geq 1$ can result in the formation of atomic collapse states (ACSs) around it (*16,17*). In previous experiments, pronounced resonances of the two types of the quasibound states were clearly observed (*9-18*). Due to their distinct underlying origins, the two quasibound states are expected to be observed in the two different systems.

Here, we demonstrate the coexistence of the electron WGMs and ACSs in graphene/WSe$_2$ heterostructure QDs. Because of the Coulomb-like potential near the edges of the QDs, we observe WGMs near the edge and ACSs in the center of the graphene/WSe$_2$ heterostructure QDs. Moreover, the ACSs are further explored in the presence of high magnetic fields. The study about effect of magnetic fields on the ACSs has a long history (*19*). However, such a longstanding prediction remains highly controversial because that the theoretical results are contradictory (*20-25*) and, more importantly, an experimental verification of this fundamental prediction is still lacking up to now (*25,26*). Our experiments demonstrate that the atomic collapse resonance effect still exist in magnetic fields and evolution from the ACSs to unusual Landau levels in the collapse regime are measured.

The graphene/WSe$_2$ heterostructure was obtained by using a wet transfer fabrication of a monolayer graphene on mechanical-exfoliated thick WSe$_2$ sheets (see methods for details of the sample preparation). In our experiment, nanoscale WSe$_2$ QDs were surprisingly observed on surface of freshly mechanical exfoliated WSe$_2$ sheets, as shown in atomic force microscopy (AFM) image of Fig. 1a. Usually, the thickness of the WSe$_2$ QDs is the same as a WSe$_2$ monolayer and the diameter is less than 10 nm. Such WSe$_2$ QDs also can be observed when mechanical-exfoliated thick WSe$_2$ is covered with graphene monolayer, *i.e.*, in the graphene/WSe$_2$ heterostructure, as shown in scanning tunneling microscope (STM) image of Fig. 1b (also see Fig. S1 for AFM images of the graphene/WSe$_2$ heterostructure). At present, the exact origin for the emergence of the nanoscale WSe$_2$ QDs is unclear. In our experiment, nanoscale monolayer-thick WSe$_2$ anti-dots are usually observed around the WSe$_2$ QDs (See Fig. S2), which suggests that the WSe$_2$ QDs are generated from the anti-dots during the process of mechanical exfoliation. Figure 1c shows an atomic-resolved STM image around a graphene/WSe$_2$ heterostructure QD. No atomic defect and strain structure (see Fig. S3 and the details of discussion) can be detected in graphene above the WSe$_2$ QD. Fast Fourier transform (FFT) images inside and outside the QD are identical (see inset of Fig. 1c and fig. S2b) and the relative rotation angle between graphene and WSe$_2$ is measured as about 20.4° (the rotation angle between the WSe$_2$ QD and the WSe$_2$ substrate is zero). A schematic side view of the graphene/WSe$_2$ heterostructure is shown in Fig. 1d.

It is interesting to find that the WSe$_2$ QD strongly modifies electronic properties of the graphene above it. Figure 1e shows two representative scanning tunneling spectroscopy (STS) spectra of the graphene on and off the WSe$_2$ QD. The STS, i.e., d*I*/d*V*, measurement of the graphene off the WSe$_2$ QD shows a typical V-shaped spectrum profile of graphene with the Dirac point at $E_D^{Off} \approx -0.2$ eV (n-doping). Whereas, spectrum of graphene on the WSe$_2$ QD displays a series of resonance peaks with the Dirac point estimated as $E_D^{On} \approx 0.3$ eV (p-doping). Obviously, the WSe$_2$ QD generates a circular p-n junction, i.e., a GQD, on graphene. The almost equally spaced peaks in the spectrum are the quasibound states confined in the GQD via the WGMs *(9-15)*. Such a result is further confirmed by carrying out STS mapping at different resonance energies (Fig. 1f and Fig. S4). For the WGMs, the quasibound states, except the lowest one, display shell structures and are progressively closer to the GQD edge with increasing the energy (Fig. S7), as observed in our experiment. In high magnetic fields, these quasibound states are condensed into Landau levels of massless Dirac fermions in graphene monolayer. These experimental results indicate that the observed peaks in the GQD are due to the confinement of the massless Dirac fermions of graphene, rather than the electronic states hosted in edges of WSe$_2$ (*27*). According to our experiment, the potential difference ($\Delta U$) on and off the GQDs shows positive correlation to the ratio ($\eta$) of the number of edge atoms to the number of inner atoms in the WSe$_2$ QD (see Fig. S5). Therefore, the large potential difference on and off the GQD may arise from the edge of the WSe$_2$ QD. The dangling bonds at the edge can significantly change the electronic property and work function of the WSe$_2$ QD (*28-30*), generating the large circular electrostatic potential on graphene covering it.

To further explore electronic properties of the GQDs, we performed the radially d$I$/d$V$ spectroscopic maps of several GQDs with different sizes and potentials, as shown in Fig. 2 (see insets of Fig. S6 for the STM images of the GQDs). By following the spatial dependence of global local density of states (LDOS) in the maps, it is interesting to note that the WSe$_2$ QDs generate a Coulomb-like electrostatic potential: $V_\beta(r) = \begin{cases} \hbar v_F \frac{\beta}{r_0}, & r \leq r_0 \\ \hbar v_F \frac{\beta}{r}, & r > r_0 \end{cases}$, where $\hbar$ is the reduced Planck constant, $v_F$ is the Fermi velocity, $r$ is the distance from the center of GQD, $\beta = Z\alpha$ with $\alpha \sim 2.5$ the fine structure constant of graphene (*8*), and $r_0$ is the cutoff radius of Coulomb potential. We obtained different values of $\beta$ and $r_0$ for different GQDs. As shown in Fig. 2, the GQDs with different $\beta$ exhibit quite different features of the quasibound states. For the $\beta = 2$ GQD, there is only one resonance peak at the center of the GQD (Fig. 2a, Top). Whereas, for the $\beta = 4.3$ GQD, besides several quasibound states confined via the WGMs at the edge of the GQD, there are three unequally spaced resonance peaks located at the center of the GQD (Fig. 2c, Top panels). To fully understand these unusual quasibound states, we numerically solved the problem for a Coulomb-like electrostatic potential with different values of $\beta$ and $r_0$ in the graphene monolayer (the values of $\beta$ and $r_0$ are extracted from our experimental results) (see supplementary information for the details). Bottom panels of Fig. 2 show the theoretical space-energy maps of the LDOS of the GQDs, which are in good agreement with the experimental results (see Fig. S6 for d$I$/d$V$ spectra and corresponding simulated LDOS at different positions of the GQDs, see Fig. S8 for the spatial distribution of the quasibound states). According to our analysis, the resonance peaks located at the edge of GQDs arise from the quasibound states via the WGMs confinement. The energy levels of these quasibound states at the edge can be estimated as $\hbar v_F / R_{eff}$ (here $R_{eff}$ is the effective radius of the GQD), as observed in our experiment (see Fig. S5d) and reported in previous studies (*10,12,13*). Whereas the energy levels of the quasibound states at the center of the GQD follow an exponential function $E_n = \frac{\hbar v_F \beta}{r_0} e^{-\frac{\pi}{\gamma}n} + E_D$, where $\gamma = \sqrt{\beta^2 - (m+1/2)^2}$, $E_D$ is the energy of Dirac point [$m$ denotes the orbital states ($m = 0, \pm1, \pm2, \ldots$)]. This is a characteristic feature of the ACSs in the supercritical regime due to the Coulomb-like electrostatic potential (see Fig. S9 and the details of discuss. The lowest quasibound state at the center of the GQD is not arising from the WGM confinement, it is the ACS, see section 10 of supplementary information for discussion) (*6-8*). Therefore, our experimental results, supported by our theoretical calculation, strongly indicate the coexistence of the WGMs and ACSs in the graphene/WSe$_2$ heterostructure GQDs. In previous studies of the Coulomb impurity in graphene with a supercritical charge, only the ACSs are observed because of the small $r_0 \sim 0.5$ nm (*16,17*). In this work, the Coulomb-like potential near the edges of the GQDs and the increase of about one order of magnitude of the $r_0$ allow us to observe both the WGMs and the ACSs.

The Coulomb-like potential also strongly affects the electronic properties of the GQDs in the presence of magnetic fields. By applying a perpendicular magnetic field, we can observe well defined Landau levels (LLs) of massless Dirac fermions at positions away from the GQD (see Fig.

S11a). When approaching the GQD, the Coulomb-like potential generates pronounced bending of the LLs (see Fig. S11a for the experimental result and theoretical simulation). Figure 3 shows radially spectroscopic maps around the $\beta = 2.4$ GQD in three different magnetic fields. Near the edge of the GQD, the bending of the LLs follows the Coulomb-like electrostatic potential. Inside the GQD, complex evolution of LDOS due to the transition from the confinement of the electrostatic potential to confinement of magnetic field is observed with increasing the magnetic field (see Fig. S12 for more experimental data). At $B = 10$ T, we can observe LLs inside the GQD. However, the $N = -1$ LL is split into three peaks: two of them with higher energies are localized in the center of the GQD and the third one is mainly located at the edge of the GQD. The splitting does not occur in pairs and the energy spacing of the splitting is as large as ~40 meV (Fig. S13), which removes valley and spin splitting as the origin of the observed phenomenon. The splitting LLs should be attributed to lifting the orbital degeneracy of LLs, which can be understood by considering the quantum-mechanical electron motion in the presence of a magnetic field and a Coulomb-like electrostatic potential. Considering the effect of the magnetic field and the electrostatic potential, the equation thus reads:

$$[v_F \vec{\sigma} \cdot (-i\hbar \vec{\nabla} + e\vec{A}) + V_\beta(\vec{r})]\psi(\vec{r}) = E\psi(\vec{r}), \qquad (1)$$

where $\vec{\sigma} = (\sigma_x, \sigma_y)$ are the Pauli matrices, $\vec{A} = (\vec{B} \times \vec{r})/2$ is the vector potential (*21,25,26*), $e$ is the electron charge. Due to the axial symmetry of the electrostatic potential in the GQD, we can describe the eigenstates by the orbital quantum number $m$ (here, we neglect spin). In the absence of the GQD, the eigen-energies $E_{Nm}$ have infinite orbital degeneracy [$\psi_{Nm}(\vec{r})$ where $m \geq -|N|$] independent of $m$ because of translational invariance. The GQD lifts this orbital degeneracy $m$ and the LLs are split into a series of sublevels, which exhibit similar behavior as that observed around charged impurities (*17,26*), due to the Coulomb-like electrostatic potential. However, previous experiments (*17,26*) in the presence of a magnetic field were limited to a charge impurity in the subcritical regime. Further, the small cutoff radius of a charge impurity prohibits to explore the evolution from the ACSs to the LLs in experiment. Such difficulties can be naturally overcome in the studied GQDs.

The detailed comparison between experiment and theory can be made by numerically solving the problem for two dimensional massless Dirac fermion of graphene monolayer in the presence of Coulomb-like electrostatic potential $V_\beta(\vec{r})$ and a magnetic field $B$ (see supplementary information for the details). The calculated radially LDOS maps in the different magnetic fields display that the orbital degeneracy is lifted, which is well consistent with our experimental results (Fig. 3). Based on the calculated results, we can identify the orbital states of the split -1 LL (Marked in Fig. 3). Thanks to the high-quality LLs in the GQD, $m = -1$ orbital state of -1 LL can be clearly identified and exhibits some characteristics distinguished from that observed in the subcritical regime (*17*,*26*). The most important feature is that the $m = -1$ orbital state can be viewed as the evolution of the ACS with increasing magnetic field. At zero magnetic field, the broad ACS is located at the center of the GQD and, interestingly, the narrower $m = -1$ orbital state appears in the same energy region in the presence of high magnetic field. Such a result indicates directly connection of the ACS and the lowest orbital state ($m = -1$) of the -1 LL.

To better explore the evolution of the ACS in the presence of magnetic fields, we summarize the measured LLs at the center of the $\beta = 2.4$ GQD as a function of the square root of the magnetic field $\sqrt{B}$ (red dots in Fig. 4, see Fig. S14a for the corresponding STS spectra). The evolution of LLs displayed a nonlinear dependence on the square root of the magnetic field, which is quite different from the feature of pristine graphene monolayer under magnetic field. The theoretical map of LDOS at the center of the $\beta = 2.4$ GQD is also plotted as a function of $\sqrt{B}$ (see supplementary information for the details, more calculated maps for different energies and magnetic fields are shown in Fig. S14), as shown in Fig. 4. With increasing magnetic field, the perturbed LLs ($N = 0$, $N = -1$, $N = -2$) display nonlinear dependence on $\sqrt{B}$. At a higher $\sqrt{B}$, the -1 LL and -2 LL are well distinctive, which split into low-energy orbital states ($m = -1$, $m = 0$). However, we did not observe the splitting of the -2 LL in the experiment, which is probably due to the large full width at half maximum (FWHM) of the LL peaks, prohibiting the observation of the splitting in the experiment. Furthermore, the ACS-R1 resonance is obvious at lower $\sqrt{B}$, and is well connected to the $m = -1$ orbital state of -1 LL (see section 14 for the detailed discussion). Similarly, ACS-R2 resonance has the similar characteristic, connected to the $m = -1$ orbital state of -2 LL. However, such a feature is harder to be recognized in the experiment due to the broadening peak of the -2 LL. Our experiments, complemented by theoretical calculations, explicitly demonstrated the existence of atomic collapse resonance effect in the presence of high magnetic fields and revealed the close connection between the ACS and the lowest orbital state ($m = -1$) of the LLs.

**References**

1. Castro Neto, A. H., Peres, N. M. R., Novoselov, K. S. & Geim, A. K. The electronic properties of graphene. *Rev. Mod. Phys.* **81**, 109-162 (2009).

2. Vozmediano, M. A. H., Katsnelson, M. I. & Guinea, F. Gauge fields in graphene. *Physics Rep.* **496**, 109-148 (2010).

3. Das Sarma, S., Adam, S., Hwang, E. & Rossi, E. Electronic transport in two-dimensional graphene. *Rev. Mod. Phys.* **83**, 407-470 (2011).

4. Goerbig, M. O. Electronic properties of graphene in a strong magnetic field. *Rev. Mod. Phys.* **83**, 1193-1243 (2011).

5. Katsnelson, M. I., Novoselov, K. S. & Geim, A. K. Chiral tunnelling and the Klein paradox in graphene. *Nat. Phys.* **2**, 620-625 (2006).

6. Pereira, V. M., Nilsson, J. & Castro Neto, A. H. Coulomb impurity problem in graphene. *Phys. Rev. Lett.* **99**, 166802 (2007).

7. Shytov, A. V., Katsnelson, M. I. & Levitov, L. S. Vacuum Polarization and Screening of Supercritical Impurities in Graphene. *Phys. Rev. Lett*. **99**, 236801 (2007).

8. Shytov, A. V., Katsnelson, M. I. & Levitov, L. S. Atomic collapse and quasi-rydberg states in graphene. *Phys. Rev. Lett*. **99**, 246802 (2007).

9. Zhao, Y. et al. Creating and probing electron whispering-gallery modes in graphene. *Science* **348**, 672-675 (2015).

10. Gutiérrez, C., Brown, L., Kim, C.-J., Park, J. & Pasupathy, A. N. Klein tunnelling and electron trapping in nanometre-scale graphene quantum dots. *Nat. Phys*. **12**, 1069-1075 (2016).

11. Lee, J. et al. Imaging electrostatically confined Dirac fermions in graphene quantum dots. *Nat. Phys*. **12**, 1032-1036 (2016).

12. Bai, K.-K. et al. Generating nanoscale and atomically-sharp p-n junctions in graphene via monolayer-vacancy-island engineering of Cu surface. *Phys. Rev. B* **97**, 045413 (2018).

13. Fu, Z. Q., Bai, K. K., Ren, Y. N., Zhou, J. J. & He, L. Coulomb interaction in quasibound states of graphene quantum dots. *Phys. Rev. B* **101**, 235310 (2020).

14. Fu, Z.-Q. et al. Relativistic Artificial Molecules Realized by Two Coupled Graphene Quantum Dots. *Nano Lett.* **20**, 6738 (2020).

15. Ghahari, F. et al. An on/off Berry phase switch in circular graphene resonators. *Science* **356**, 845 (2017).

16. Wang, Y. et al. Observing atomic collapse resonances in artificial nuclei on graphene. *Science* **340**, 734-737 (2013).

17. Mao, J. et al. Realization of a tunable artificial atom at a supercritically charged vacancy in graphene. *Nat. Phys*. **12**, 545-549 (2016).

18. Jiang, Y. et al. Tuning a circular p-n junction in graphene from quantum confinement to optical guiding. *Nat. Nanotechnol*. **12**, 1045-1049 (2017).

19. Reinhardt, J. & Greiner, W. Quantum electrodynamics of strong fields. *Rep. Prog. Phys.* **40**, 219-295 (1977).

20. Gamayun, O. V., Gorbar, E. V. & Gusynin, V. P. Magnetic field driven instability of a charged center in graphene. *Phys. Rev. B* **83**, 235104 (2011).

21. Sobol, O. O., Pyatkovskiy, P. K., Gorbar, E. V. & Gusynin, V. P. Screening of a charged impurity in graphene in a magnetic field. *Phys. Rev. B* **94**, 115409 (2016).

22. Zhang, Y., Barlas, Y. & Yang, K. Coulomb impurity under magnetic field in graphene: A semiclassical approach. *Phys. Rev. B* **85**, 165423 (2012).

23. Maier, T. & Siedentop, H. Stability of impurities with Coulomb potential in graphene with homogeneous magnetic field. *J. Math. Phys.* **53**, 095207 (2012).

24. Kim, S. C. & Eric Yang, S. R. Coulomb impurity problem of graphene in magnetic fields. *Ann. Phys.* **347**, 21-31 (2014).

25. Moldovan, D., Masir, M. R. & Peeters, F. M. Magnetic field dependence of the atomic collapse state in graphene. *2D Mater.* **5**, 015017 (2018).

26. Luican-Mayer, A. et al. Screening charged impurities and lifting the orbital degeneracy in graphene by populating landau levels. *Phys. Rev. Lett.* **112**, 036804 (2014).

27. Zhang, C. et al. Visualizing band offsets and edge states in bilayer–monolayer transition metal dichalcogenides lateral heterojunction. *Nat. Commun.* **7**, 10349 (2016).

28. Zhang, Y. et al. Electronic Structure, Surface Doping, and Optical Response in Epitaxial $WSe_2$ Thin Films. *Nano Lett.* **16**, 2485-2491 (2016).

29. Addou, R. & Wallace, R. M. Surface Analysis of $WSe_2$ Crystals: Spatial and Electronic Variability. *ACS Appl. Mater. Interfaces* **8**, 26400-26406 (2016).

30. Kahn, A. Fermi level, work function and vacuum level. *Mater. Horiz.* **3**, 7-10 (2016).

**Acknowledgments:**

This work was supported by the National Natural Science Foundation of China (Grant Nos. 11974050, 11674029, 11921005) and National Key R and D Program of China (Grant No. 2017YFA0303301). L.H. also acknowledges support from the National Program for Support of Top-notch Young Professionals, support from "the Fundamental Research Funds for the Central Universities", and support from "Chang Jiang Scholars Program".

**Author contributions**

Q.Z. performed the sample synthesis, characterization and STM/STS measurements. Q.Z., Y.C.Z., and L.H. analyzed the data. Y.C.Z. carried out the theoretical calculations. L.H. conceived and provided advice on the experiment and analysis. Q.F.S. conceived and provided advice on the theoretical calculations. Q.Z. and L.H. wrote the paper with the input from others. All authors participated in the data discussion.

**Data availability statement**

All data supporting the findings of this study are available from the corresponding author upon request.

**Methods**

**CVD Growth of Graphene.** The large area graphene monolayer films were grown on a $20\times 20 \text{ mm}^2$ polycrystalline copper (Cu) foil (Alfa Aesar, 99.8% purity, 25 μm thick) via a low pressure chemical vapor deposition (LPCVD) method. The cleaned Cu foil was loaded into one quartz boat in center of the tube furnace. Ar flow of 50 sccm (Standard Cubic Centimeter per Minutes) and $H_2$ flow of 50 sccm were maintained throughout the whole growth process. The Cu foil was heated from room temperature to 1030 ℃ in 30 min and annealed at 1030 ℃ for six hours. Then $CH_4$ flow of 5 sccm was introduced for 20 min to grow high-quality large area graphene monolayer. Finally, the furnace was cooled down naturally to room temperature.

**Construction of graphene/WSe₂ heterostructure.** We used conventional wet etching technique with polymethyl methacrylate (PMMA) to transfer graphene monolayer onto the substrate. PMMA was first uniformly coated on Cu foil with graphene monolayer. We transferred the Cu/graphene/PMMA film into ammonium persulfate solution, and then the underlying Cu foil was etched away. The graphene/PMMA film was cleaned in deionized water for hours. The WSe$_2$ crystal was separated into thick-layer WSe$_2$ sheets by traditional mechanical exfoliation technology and then transferred to $8\times 8$ mm$^2$ highly N-doped Si wafer [(100) oriented, 500 μm thick]. We placed graphene/PMMA onto Si wafer which has been transferred with WSe$_2$ sheets in advance. Finally, the PMMA was removed by acetone and then annealed in low pressure with Ar flow of 50 sccm and H$_2$ flow of 50 sccm at ~300 ℃ for 1 hours.

**AFM, STM and STS Measurements.** The topographical images are measured by atomic force microscope (AFM, Bruker Multimode 8) with a tapping mode. We employed the n-doped Si tip coated with Platinum-Iridium (Bruker, SCM-PIT-V2, frequency 50-100KHz, spring constant 1.5-6 N/m) to characterize WSe$_2$ and graphene/WSe$_2$ heterostructure samples. STM/STS measurements were performed in low-temperature (77 K for Fig. S5a and c, 4.2 K for Fig. S5b) and ultrahigh-vacuum (~10$^{-10}$ Torr) scanning probe microscopes [USM-1400 (77 K) and USM-1300 (4.2 K)] from UNISOKU. The tips were obtained by chemical etching from a Pt/Ir (80:20%) alloy wire. The differential conductance (d$I$/d$V$) measurements were taken by a standard lock-in technique with an ac bias modulation of 5 mV and 793 Hz signal added to the tunneling bias.

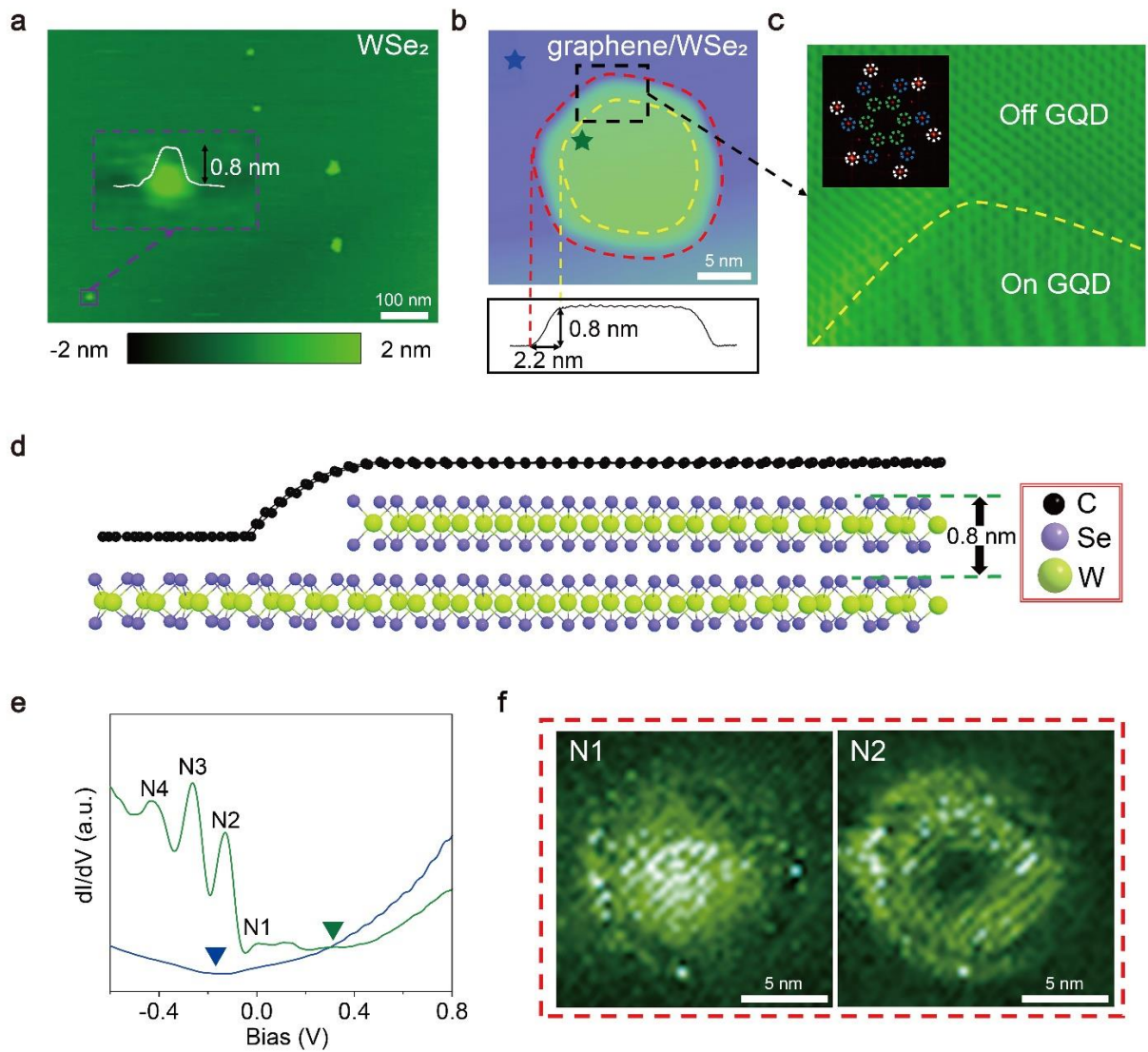

**Fig. 1 | The structures and general d$I$/d$V$ features of Graphene/WSe$_2$ heterostructure. a**, A representative AFM image of the freshly mechanical exfoliated WSe$_2$ sheet. Inset: the AFM image of a typical monolayer WSe$_2$ island. **b**, A STM image of a typical graphene/WSe$_2$ heterostructure QD. The height of the GQD is ~0.8 nm and the width of edge area of the GQD is ~2.2 nm. **c**, The zoom-in image of the area in black dashed squares from panel **b**. Inset: the FFT of graphene/WSe$_2$ heterostructure. The bright spots in the white dotted circles represent the reciprocal lattice of graphene, the bright spots in the blue dotted circles represent the reciprocal lattice of WSe$_2$, and the bright spots in the green dotted circles represent moiré structure of the graphene/WSe$_2$

heterostructure. **d**, Schematic structure of the graphene/WSe$_2$ heterostructure QD. **e**, The d*I*/d*V* spectra taken inside [marked by dark green pentagram in **b**] and outside [marked by blue pentagram in **b**] the GQD. **f**, The d*I*/d*V* maps with different energies [N1 and N2 marked in **e**] of the GQD.

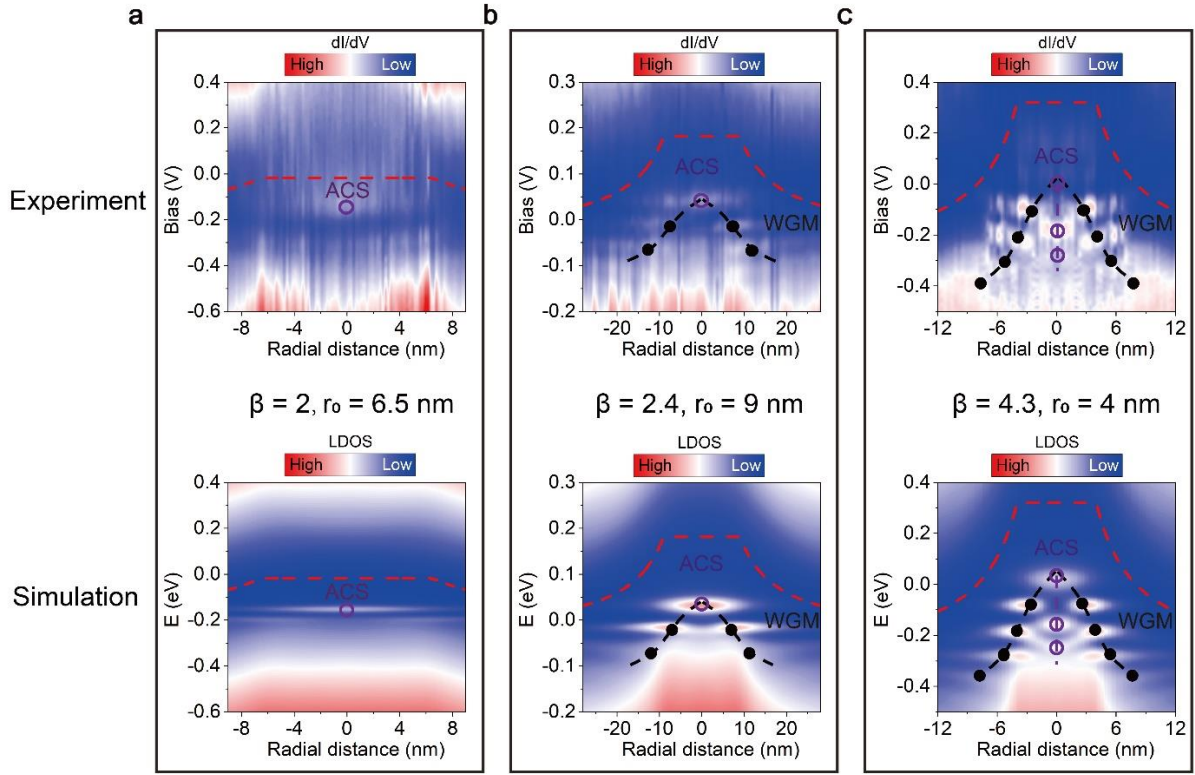

**Fig. 2 | Coexistence of WGMs confinement and ACSs in the GQDs.** Top of **a** to **c**, The radially d*I*/d*V* spectroscopic maps of different GQDs. Bottom of **a** to **c**, The calculated space-energy maps of the LDOS of different GQDs with different value of $\beta$ and $r_0$. The red dotted lines indicate Dirac point energy. The black solid dots indicate the quasibound states via the WGM confinement, and the purple hollow dots indicate the ACSs.

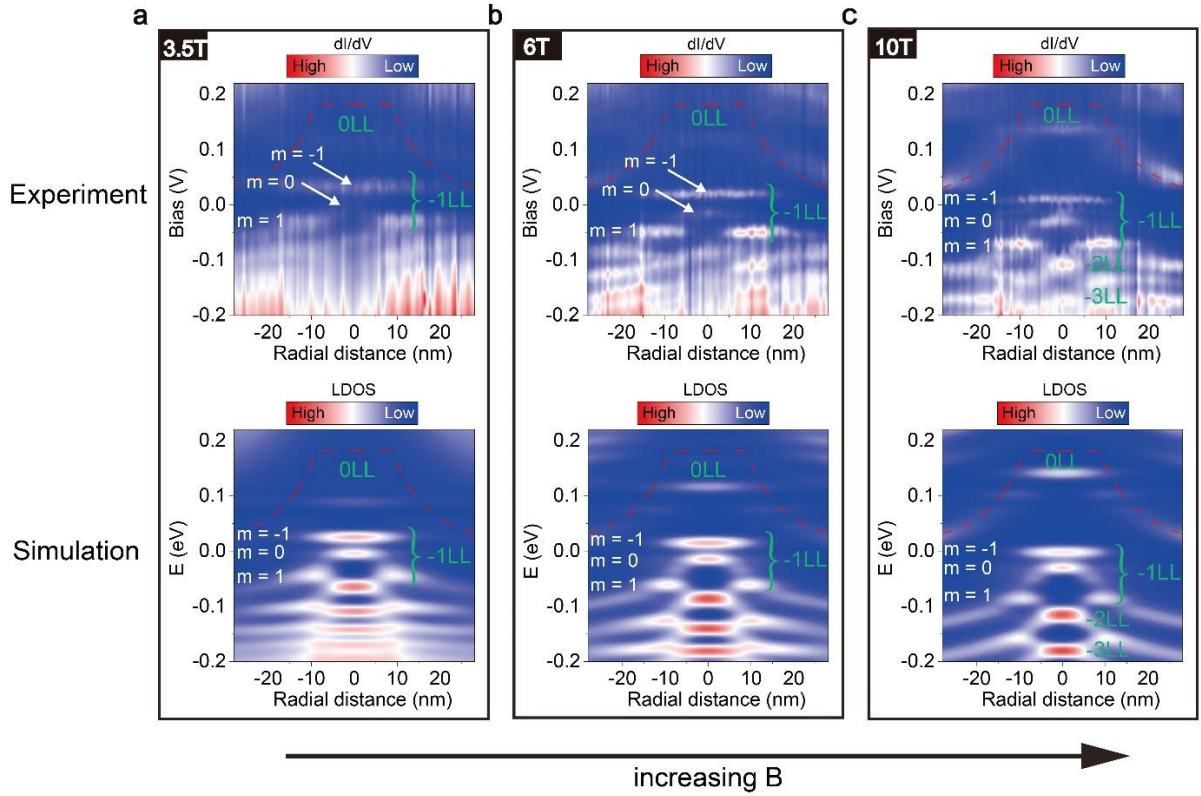

**Fig. 3 | Lifting the orbital degeneracy of LLs in the GQD.** Top panels of **a** to **c**, the radially d$I$/d$V$ spectroscopic maps on the GQD ($\beta = 2.4$, $r_0 = 9$ nm) in the case of a series of magnetic fields. Bottom panels of **a** to **c**, the calculated space-energy maps of the LDOS of the GQD with different magnetic fields. The $m = -1$, $m = 0$, and $m = 1$ indicate the split orbital states of the -1 LL. The red dotted lines indicate Dirac point energy.

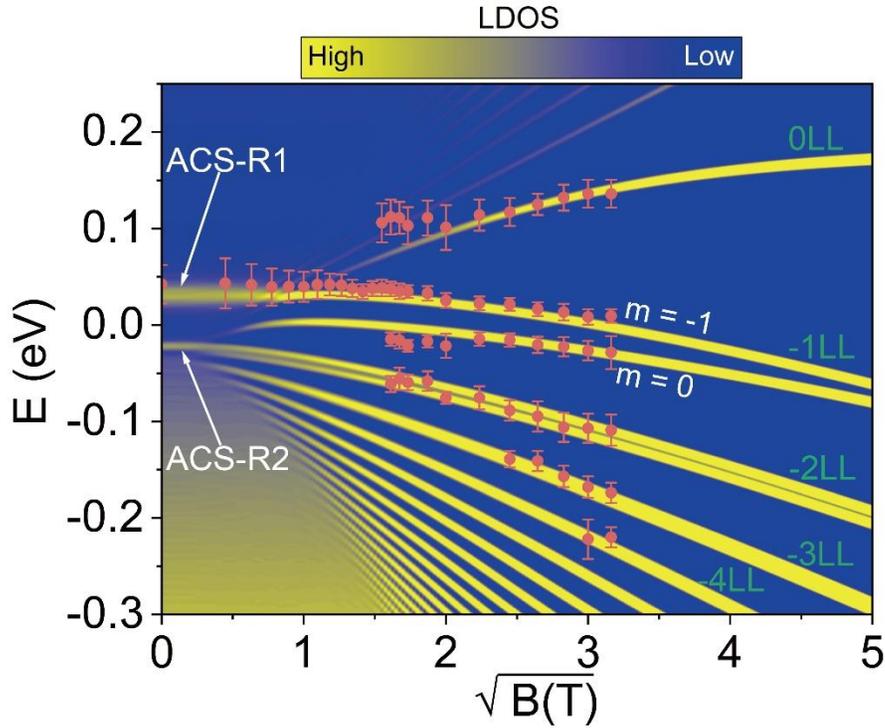

**Fig. 4 | Evolution from the ACSs** to unusual LLs in the GQD. The measured LLs at the center of the GQD ($\beta = 2.4$, $r_0 = 9$ nm) as a function of the square root of the magnetic field $\sqrt{B}$. The experimental results are superimposed onto the calculated map of LDOS in the GQD with $\beta = 2.4$ and $r_0 = 9$ nm. The ACS-R1 and ACS-R2 are two quasi-bound states due to atomic collapse resonance. The full width at half maximum of the peaks in the spectra was used to estimate the error bar in experiment (orange dots).